\documentclass[lettersize,journal]{IEEEtran}
\usepackage{amsmath,amsfonts}
\usepackage{algorithmic}
\usepackage{algorithm}
\usepackage{array}
\usepackage[caption=false,font=normalsize,labelfont=sf,textfont=sf]{subfig}
\usepackage{textcomp}
\usepackage{stfloats}
\usepackage{url}
\usepackage{verbatim}
\usepackage{graphicx}
\usepackage{cite}
\usepackage{booktabs}   % 用于 \toprule, \midrule, \bottomrule
\usepackage{amssymb}    % 用于 \checkmark
\usepackage{amsmath}    % 用于 \mathcal
\usepackage{multirow}
\usepackage{graphicx}   % 用于调整表格宽度

\usepackage{siunitx}
\usepackage{amsmath,amssymb,amsfonts}
\usepackage{wrapfig}
\usepackage{threeparttable}

\begin{document}

\title{Physically Motivated Knowledge Distillation for Blind Geometric Correction of Side-Scan Sonar Imagery}

\author{Can Lei,
        Hayat Rajani,~\IEEEmembership{Member,~IEEE},
        Valerio Franchi,
        Rafael Garcia,
        Nuno Gracias,       
        Huigang Wang,~\IEEEmembership{Member,~IEEE},
        and Wei Qiang
        % <-this % stops a space
\thanks{Can Lei and Huigang Wang are with the School of Marine Science and Technology, Northwestern Polytechnical University, Xi’an 710072, China. Can Lei and Wei Qiang is with the Advanced Institute for Ocean Research, Southern University of Science and Technology, Shenzhen, 518055, China. Huigang Wang is also with the Research \& Development Institute of Northwestern Polytechnical University in Shenzhen, Shenzhen 518057, China.}% <-this % stops a space

\thanks{Hayat Rajani, Valerio Franchi, Nuno Gracias and Rafael Garcia are with the Computer Vision and Robotics Research Institute (ViCOROB) of the University of Girona, Spain. This work was conducted while Can Lei was on a research stay at ViCOROB, Spain}% <-this % stops a space

\thanks{This work was partly supported by the Spanish government through projects ASSiST (PID2023-149413OB-I00) and IURBI (CNS2023-144688). This work was also supported by the Science, Technology and Innovation of Shenzhen Municipality (JCYJ20241202124931042, ZDCYKCX20250901093900002).} % <-this % stops a space
\thanks{Corresponding author: Huigang Wang (e-mail: wanghg74@nwpu.edu.cn).}}

\maketitle

\begin{abstract}

Side-scan sonar (SSS) imagery is susceptible to geometric distortions caused by platform motion instability, which degrade geometric consistency and limit downstream analyses such as mosaicking and perception. Conventional correction methods typically rely on navigation and attitude measurements, which are often unreliable in real ocean conditions. This unreliability necessitates blind geometric correction from a single distorted image, a highly ill-posed problem. To address this issue, we propose a physically motivated knowledge distillation framework for blind geometric correction of SSS imagery. Specifically, a teacher network is trained using paired distorted and geocoded reference images to learn distortion-related geometric differences, and this knowledge is transferred to a student network that performs correction using only a single distorted image during blind inference. To ensure physically plausible deformation estimation, we design a parametric decoder that represents distortions as row-wise affine transformations consistent with the SSS line-scanning imaging mechanism. To compensate for the absence of reference information during blind inference, a hallucination context module is introduced to approximate the teacher’s geometric reasoning from distorted features under a multi-level distillation scheme. In addition, a differentiable forward warping strategy is adopted to handle the non-bijective deformation characteristics of SSS imagery in an end-to-end manner. Extensive experiments on multiple datasets show that the proposed method outperforms state-of-the-art baselines and generalizes well across different platforms and acquisition conditions.
\end{abstract}


\section{Introduction}
\IEEEPARstart{S}{ide}-scan sonar (SSS) is a primary instrument in underwater remote sensing, essential for seabed mapping \cite{A1} and target detection \cite{A2}. Unlike frame-based optical cameras, SSS adopts a line-scanning imaging mechanism, where acoustic backscatter measurements are acquired sequentially along the platform trajectory and stacked into a two-dimensional image \cite{A3}.  While enabling wide swath coverage, this process makes SSS highly sensitive to platform instability, leading to severe geometric distortions.

Ideally, the sonar platform follows a straight trajectory with constant velocity and stable attitude. However, ocean currents, vehicle control errors, and mechanical disturbances inevitably cause motion variations. These instabilities induce geometric distortions in SSS imagery, where velocity and pitch variations lead to compression or stretching along the along-track direction, while yaw causes sinuous and rotational distortions \cite{A4}. Furthermore, coupled motions introduce complex, range-dependent deformations, severely degrading geometric consistency and hindering downstream interpretation.

From the imaging perspective, an SSS image represents a 2D projection of the 3D seabed acquired under dynamic platform motion. Conventional geometric correction relies on sensor-based geocoding \cite{A5}, which requires accurate and synchronized auxiliary measurements from GPS, Doppler velocity logs, and inertial measurement units to correct distortions through projection models,  such as the Universal Transverse Mercator framework. However, in scenarios involving low-cost autonomous underwater vehicles, or cases with sensor noise, drift, or synchronization errors, reliable auxiliary data is absent. In these situations, analysts are restricted to decoded waterfall images, rendering sensor-based approaches infeasible. Therefore, performing geometric correction only from distorted SSS imagery becomes crucial but extremely challenging. Without motion priors, recovering plausible deformation fields is inherently ambiguous. While recent deep learning–based geometric correction methods, mostly developed for other imaging domains, are typically formulated as generic image-to-image transformation problems \cite{A6}, thereby failing to capture the line-scanning geometry and motion constraints specific to SSS imagery.

To address these challenges, we propose a physically motivated knowledge distillation framework for blind SSS geometric correction. Utilizing a teacher–student architecture, the teacher learns from paired data to encode privileged features characterizing motion-induced distortions, transferring this privileged knowledge to a student network that infers from single distorted images. A context hallucination module is integrated to infer geometric correction information, allowing the student to approximate the teacher’s geometry-aware reasoning process. Furthermore, a shared parametric decoder models distortions as row-wise affine transformations consistent with the SSS line-scanning mechanism, ensuring physically plausible estimation during blind inference. The main contributions of this paper are summarized as follows:

\begin{itemize}
    \item We propose a teacher–student physically motivated knowledge distillation framework for blind SSS geometric correction, transferring privileged knowledge from paired distorted and geocoded images to a student network operating on single distorted inputs.

    \item We introduce a parametric decoder that models distortions as row-wise affine transformations consistent with the SSS line-scanning mechanism, ensuring the physical plausibility of estimated deformation fields.

    \item We design a hallucination context module coupled with a multi-level distillation strategy, enabling the student network to infer geometric correction information using only distorted inputs during blind inference.

    \item We incorporate a differentiable forward warping strategy combining soft splatting and diffusion-based filling to handle the non-bijective deformation characteristics of SSS imagery and enable end-to-end optimization.

\end{itemize}

\section{Related Work}

\subsection{Traditional Geometric Correction for Side-Scan Sonar}

Traditional geometric correction methods for SSS imagery can be broadly grouped into image-driven approaches and geocoding-based approaches. Image-driven methods estimate motion-induced distortions directly from image statistics, without relying on navigation or attitude measurements. A representative work by Cobra et al. \cite{A7} estimates local geometric distortion through cross-correlation of adjacent scanlines, fits a physically motivated sampling model under a planar seabed assumption, and resamples the image to improve geometric consistency.  Following this work, relatively little research has focused on purely image-driven geometric correction, mainly due to its strong assumptions on seabed geometry and image statistics, which limit robustness in complex scenes.

Geocoding-based methods correct geometric distortions by modeling the sonar acquisition geometry using onboard navigation and attitude data \cite{A81}. Sheffer et al. \cite{A8} proposed a unified correction pipeline that accounts for slant range, platform speed, position, yaw, and pitch, mapping each sonar ping to a consistent geometric location under a flat-seabed assumption. Ye et al. \cite{A9} extended this framework by incorporating altitude information to handle terrain-related range variations and by using GPS data to correct along-track distortions, enabling more accurate reconstruction over uneven seabed topography. More recently, Franchi et al. \cite{A10} further advanced this approach by performing scanline-level geometric remapping with yaw, pitch, and altitude measurements, projecting sonar intensities into a georeferenced mosaic and reconstructing distortion-reduced waterfall images through interpolation and trajectory smoothing.

Overall, geocoding-based methods can achieve geometrically accurate correction when high-quality auxiliary data are available, but their performance strongly depends on the accuracy and reliability of such measurements. Image-driven methods, on the other hand, remove this dependency and are therefore suitable for blind correction settings, but their effectiveness is constrained by simplified geometric and statistical assumptions. These limitations indicate that traditional blind geometric correction models is insufficient, motivating the development of learning-based approaches that estimate geometric distortions directly from sonar imagery without external measurements.

\subsection{Deep Learning-based Geometric Correction}

\begin{figure*}
	\centering
	\includegraphics[width=1\textwidth]{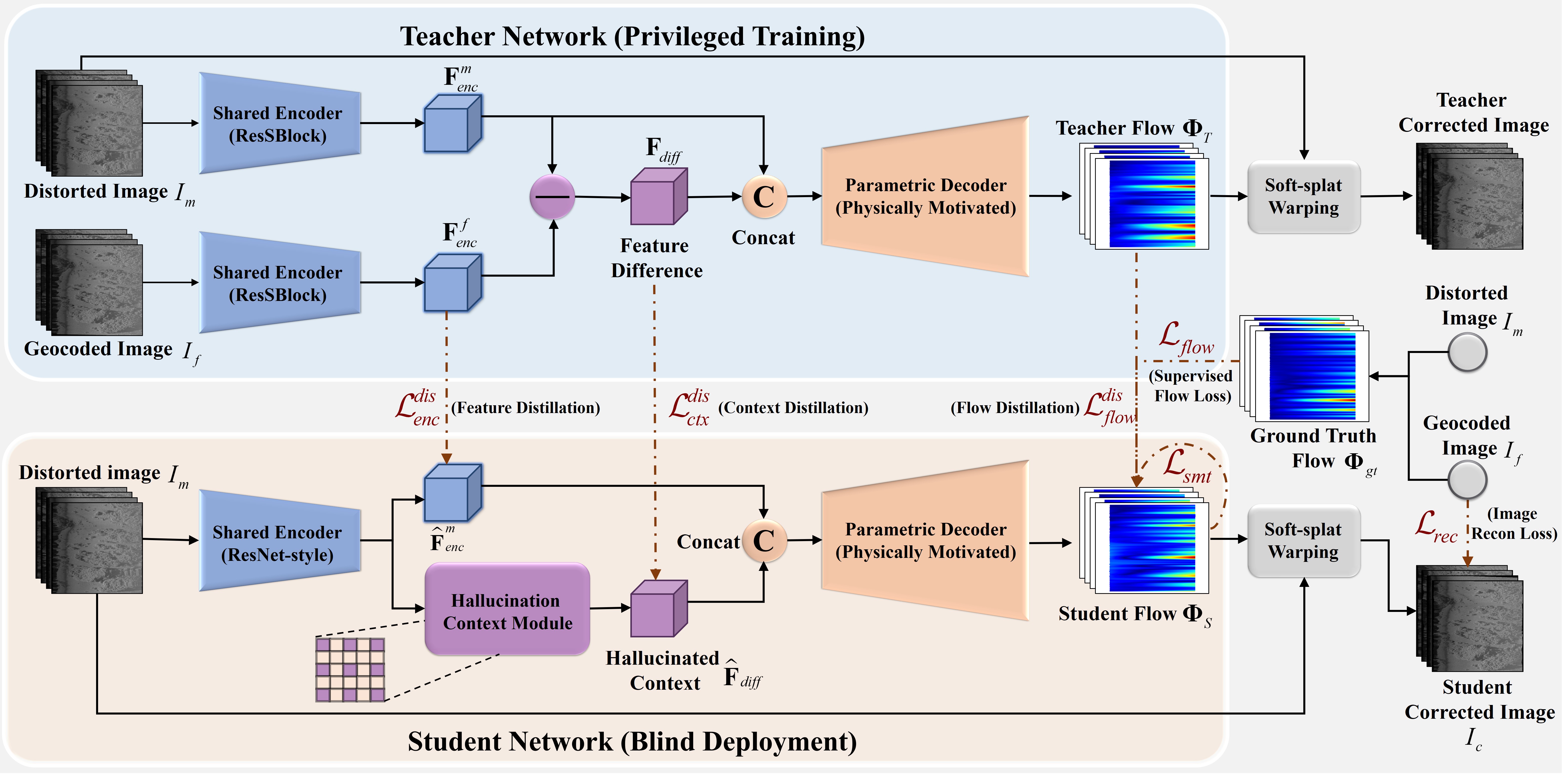}
	\caption{Overview of the proposed physically motivated knowledge distillation framework for blind geometric correction of side-scan sonar imagery. The Teacher network is fed with paired distorted and geocoded images $\{I_m, I_f\}$ to explicitly extract geometric difference features and estimate a physically motivated deformation flow. The Student network operates under blind conditions using only distorted input $I_m$, where a hallucination context module approximates the missing geometric difference and is guided by multi-level distillation from the Teacher. Both networks share a physically motivated parametric decoder and employ differentiable forward warping to obtain the corrected image.}
	\label{fig1}
\end{figure*}	

While deep learning has been widely applied to SSS imagery for detection \cite{A11}, segmentation \cite{A12}, and classification tasks \cite{A13}, its use for blind geometric correction remains largely unexplored. We therefore draw methodological insights from deep learning-based geometric correction studies in computer vision, which mainly include rolling shutter (RS) correction, lens distortion correction, and document image rectification.

RS correction is closely related to SSS imaging, as both suffer from motion-induced distortions caused by platform movement coupled with row-wise acquisition \cite{A14}. Fan et al. \cite{A15} established an explicit linear scaling mapping from optical flow to RS correction flow under a constant motion assumption, while Cao et al. \cite{A16} improved robustness under large motions by introducing global relevance attention. To avoid simplified motion assumptions, Erbach et al. \cite{A17} leveraged event-based measurements and a Filter-and-Flip transformation for single-frame correction of complex nonlinear distortions. Despite their effectiveness, RS correction methods depend on reliable texture-based motion estimation, which is often unreliable in weak-texture sonar imagery. Lens distortion correction, particularly fisheye rectification, addresses geometric artifacts from non-ideal optical projection \cite{A18}. Li et al. \cite{A19} proposed a CNN-based correction framework, Yang et al. \cite{A20} incorporated straight-line priors for unsupervised correction, and Chen et al. \cite{A21} and Xu et al. \cite{A22} further enhanced performance under severe radial distortion using joint texture and edge modeling and Transformer-based architectures, respectively. However, these methods rely on linear or globally consistent geometric structures, which are rarely present in unstructured seabed terrain, limiting their applicability to sonar imagery.

Document image correction targets non-rigid distortions caused by perspective projection of non-planar documents \cite{A23}. Ma et al. \cite{A24} proposed DocUNet to predict dense displacement fields using stacked U-Net architectures, demonstrating the feasibility of learning dense geometric mappings. To capture long-range dependencies, , Liao et al. \cite{A25} introduced DocTr based on deformable attention, while Feng et al. \cite{A26} further integrated hierarchical encoder–decoder designs with self-attention for robust correction. Although these methods highlight the importance of global context modeling, they rely on clear structures and near-bijective mappings, assumptions that are incompatible with the noise, and information loss inherent in sonar imaging.

In summary, despite targeting different distortion sources, existing deep learning-based geometric correction methods rely on learning dense deformation fields from appearance information, with additional motion assumptions or architectural constraints. This dependence on appearance statistics makes them vulnerable to physically inconsistent deformations when correspondences are unreliable or mappings are non-bijective, conditions that are common in sonar imagery. These limitations motivate our approach, which departs from purely data-driven learning by embedding physical constraints and privileged geometric knowledge for robust blind correction.

\section{Methodology}

\subsection{Problem Formulation and Overview}

The objective of blind geometric correction for SSS imagery is to recover a geographically corrected image $I_c$ from a distorted observation $I_m \in \mathbb{R}^{H \times W}$, without relying on external navigation or attitude measurements. Unlike frame-based optical imaging, an SSS waterfall image is formed sequentially along the platform trajectory, so the geometric position of each scanline depends on the platform state at its acquisition time. As a result, motion instability introduces scanline-dependent distortions, such as compression, stretching, bending, and local rotation \cite{A7}. Since adjacent scanlines are acquired under similar poses, these distortions exhibit strong row-wise coherence rather than arbitrary dense deformation. We therefore formulate blind SSS correction using a row-dependent parametric deformation model.

Formally, we aim to learn a mapping function $\mathcal{F}$ that estimates a deformation flow field $\mathbf{\Phi}$, which models the geometric distortions caused by the platform's motion instabilities. The corrected image is then obtained by warping $I_m$ through the predicted flow field:
\begin{equation}
    I_c = \mathcal{W}(I_m, \mathbf{\Phi}) \approx I_f, \quad\mathbf{\Phi} = \mathcal{F}(I_m)
\end{equation}
where $\mathcal{W}(\cdot )$ denotes the differentiable warping operator, and $I_f$ denotes the geocoded image reconstructed from high-precision navigation and attitude measurements.

However, inferring $\mathbf{\Phi}$ from the single distorted observation $I_m$ is a severely ill-posed inverse problem due to the lack of explicit motion parameters at inference time. To address this, we depart from generic image-to-image transformation and propose a physically motivated knowledge distillation framework. As illustrated in Fig. \ref{fig1}, our method adopts a Teacher-Student paradigm designed to transfer geometric knowledge from a privileged training environment to a blind inference model:

\begin{itemize}
    \item \textbf{Teacher Network (Privileged Training):} During training, the Teacher network is fed with paired samples $\{I_m, I_f\}$. By encoding both the distorted input and the rectified reference, the Teacher explicitly computes a feature difference $\mathbf{F}_{diff}$, which captures the geometric difference context between pairs and serves as a supervision for blind correction.
    
    \item \textbf{Student Network (Blind Deployment):} The Student network is designed for real-world scenarios where only $I_m$ is available. To compensate for the missing reference $I_f$, a Hallucination Context Module (HCM) is introduced to predict an approximate feature difference $\hat{\mathbf{F}}_{diff}$ from the distorted input, while a multi-level distillation strategy enforces consistency with the Teacher across encoder features, geometric difference, and predicted flow. This joint design enables the Student to capture the Teacher’s geometric inference behavior under blind conditions.
\end{itemize}

To ensure physical plausibility in deformation estimation, both networks employ a shared physically motivated parametric decoder that restricts the deformation to row-wise affine transformations, consistent with the line-scanning imaging mechanism of SSS. Based on the resulting deformation field $\mathbf{\Phi}$, the corrected image is obtained through a Differentiable Forward Warping module, which enables end-to-end optimization of the proposed framework.

\subsection{Physically Motivated Deformation  Estimation}

Unlike generic optical flow, the deformation in SSS imagery is induced by the interaction between platform motion and line-scanning acquisition. As discussed previously, the resulting distortion field exhibits strong scanline-wise coherence and can be more appropriately described by a row-dependent low-dimensional model rather than a fully unconstrained dense displacement field. Based on this observation, we design a physically motivated encoder-decoder architecture for deformation estimation, where the encoder extracts geometry-relevant features and the decoder enforces the scanline-structured deformation prior.

\subsubsection{Geometry-relevant Feature Encoder}

\begin{figure}
	\centering
	\includegraphics[width=1\columnwidth]{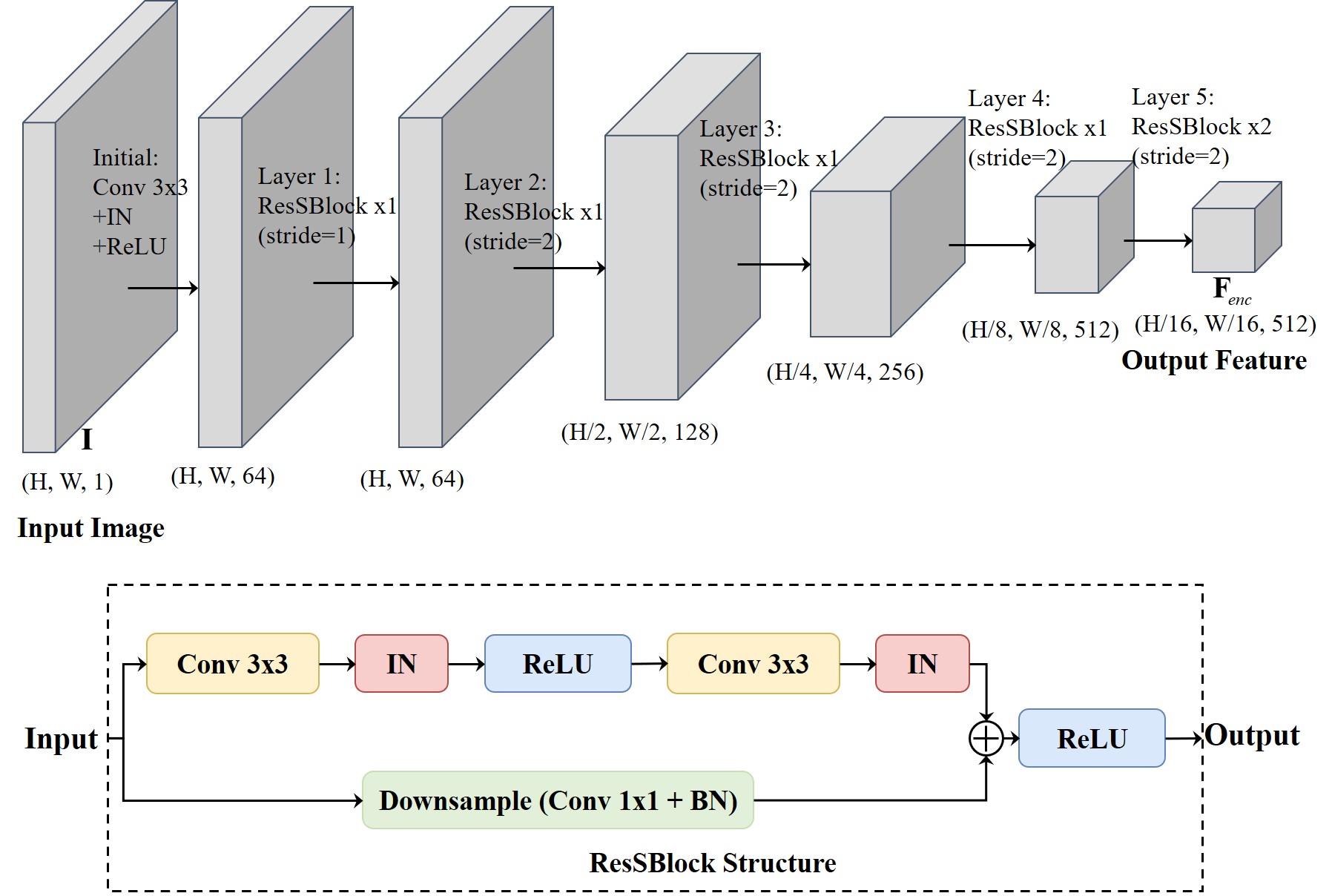}
	\caption{Structure of the shared encoder used in both Teacher and Student networks. An initial $3\times3$ convolution with Instance Normalization (IN) projects the input image into a 64-channel feature space, followed by five cascaded Sonar-Residual Blocks (ResSBlock). The first stage preserves spatial resolution, while the remaining stages progressively downsample the feature maps, producing a compact and noise-robust representation $\mathbf{F}_{enc}$ for subsequent geometric inference.}
	\label{fig2}
\end{figure}

Backscatter intensity in SSS imagery is highly sensitive to grazing angle variations and seabed properties, resulting in radiometric changes that are largely independent of geometric distortions. To decouple such variations from geometric information, we employ a shared encoder $E(\cdot)$ with Instance Normalization (IN) \cite{A27}, which suppresses sample-specific intensity bias. Accordingly, the encoder is designed for stable and noise-robust feature learning, while geometric constraints are explicitly introduced in the parametric decoder, where physical priors can be naturally enforced. As illustrated in Fig. \ref{fig2}, the encoder processes the input image $\mathbf{I} \in \mathbb{R}^{H \times W \times 1}$ through a hierarchical residual architecture and produces a compact latent representation:
\begin{equation}
\mathbf{F}_{enc} = E(\mathbf{I}) \in \mathbb{R}^{\frac{H}{16} \times \frac{W}{16} \times 512}.
\end{equation}

Specifically, the encoding pipeline begins with a $3\times3$ convolutional projection layer that maps $\mathbf{I}$ into a 64-channel feature space, followed by five cascaded Sonar-Residual Block (ResSBlock) stages. The first stage preserves the original spatial resolution to retain fine-grained texture, while the remaining stages progressively downsample the feature maps using stride-2 convolutions. This staged downsampling process transforms local texture into abstract and context-aware representations, providing a stable and noise-robust feature for subsequent geometric inference in the decoder.

In the Teacher network, the encoder processes $I_m$ and $I_f$ through two weight-sharing branches, producing the encoded features $\mathbf{F}_{enc}^{m}$ and $\mathbf{F}_{enc}^{f}$, respectively, while the Student uses a single branch to process $I_m$, yielding the feature $\hat{\mathbf{F}}_{enc}^{m}$.

\subsubsection{Parametric Decoder}

To embed the structural regularities of SSS distortions into the geometric correction process, we propose a physically motivated parametric decoder that models image distortion in a row-wise parametric form, where each scan line is governed by a low-dimensional affine transformation, thereby substantially reducing the solution space.

\begin{figure}
	\centering
	\includegraphics[width=1\columnwidth]{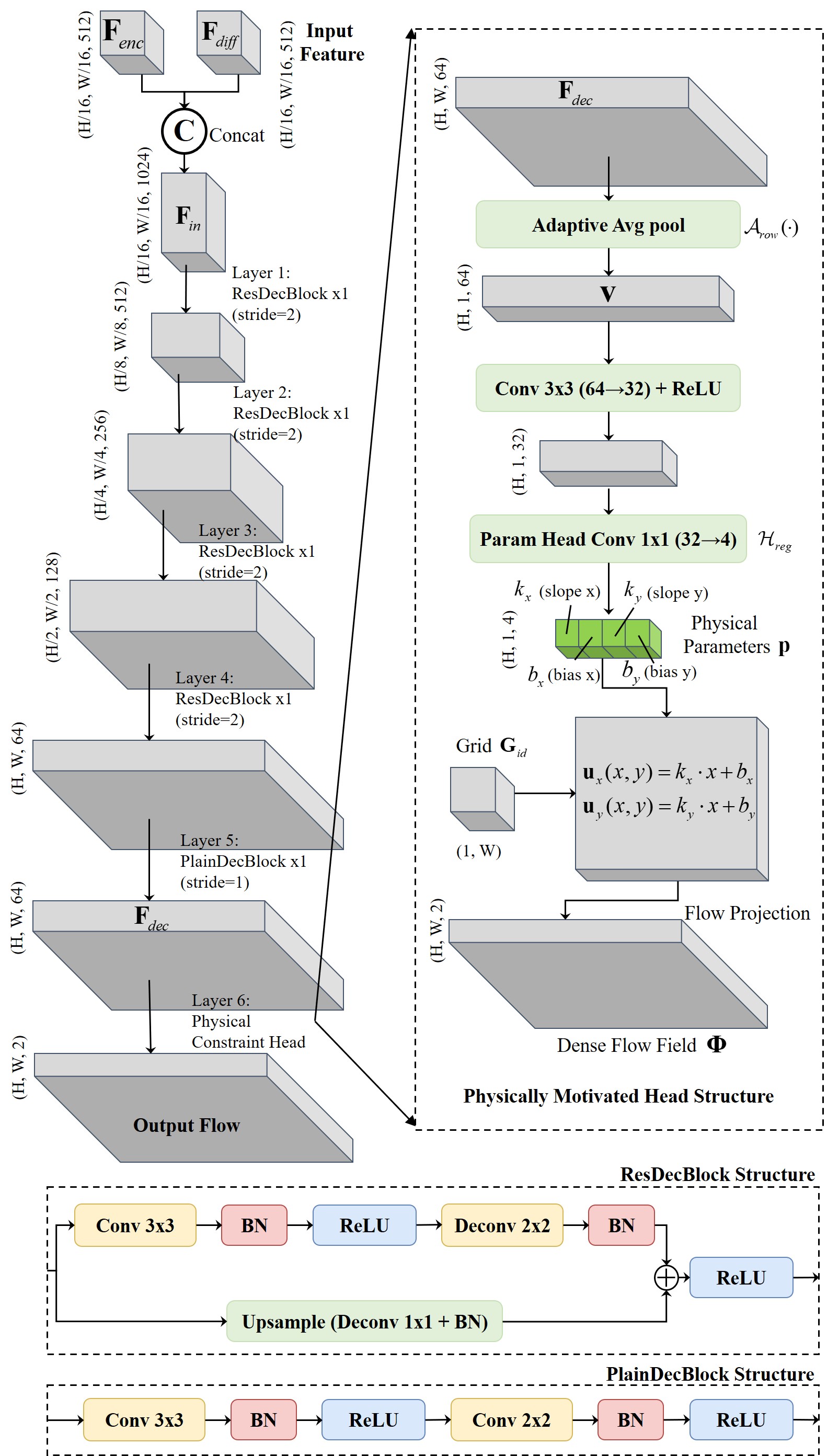}
	\caption{Structure of the physically motivated parametric decoder. Encoder features $\mathbf{F}_{enc}$ and difference features $\mathbf{F}_{diff}$ are concatenated and progressively upsampled through a hierarchical decoder composed of Residual Decoding Blocks (ResDecBlock) and a Plain Decoding Block (PlainDecBlock), producing a dense feature map $\mathbf{F}_{dec}$. A physically motivated head then aggregates features along each scan line and regresses row-wise affine parameters, which are projected into a dense deformation flow field $\mathbf{\Phi}$ consistent with the line-scanning geometry of side-scan sonar imagery.}
	\label{fig4}
\end{figure}	

\paragraph{Hierarchical Feature Decoding} 

The decoding process begins with the fused latent representation produced by the encoder and the difference-aware module. Specifically, the encoder features $\mathbf{F}_{enc}$ and the difference features $\mathbf{F}_{diff}$ (or $\hat{\mathbf{F}}_{diff}$ in the student network) are concatenated along the channel dimension to form a bottleneck tensor describing global structure and geometric discrepancies:
\begin{equation} 
    \mathbf{F}_{in} = \text{Concat}(\mathbf{F}_{enc}, {\mathbf{F}}_{diff}) \in \mathbb{R}^{ \frac{H}{16}\times \frac{W}{16} \times 1024}. 
\end{equation}

The decoder backbone $\mathcal{D}(\cdot)$ then progressively restores spatial resolution through a hierarchical upsampling process. As illustrated in Fig. \ref{fig4}, $\mathcal{D}(\cdot)$ consists of four cascaded Residual Decoding Blocks (ResDecBlock), each combining learned upsampling with convolutional refinement, followed by a Plain Decoding Block (PlainDecBlock) to stabilize the final feature representation. Through this staged decoding strategy, high-level features are gradually propagated to the pixel domain, generating a dense feature map:
\begin{equation}
    \mathbf{F}_{dec} = \mathcal{D}(\mathbf{F}_{in})
    \in \mathbb{R}^{H \times W \times 64}.
\end{equation}

\paragraph{Physically Motivated Head} 

To translate $\mathbf{F}_{dec}$ into physically plausible geometric deformations, we exploit the row-wise consistency of the SSS line-scanning mechanism by aggregating features along each scanline to estimate a set of row-dependent transformation parameters. Specifically, we introduce a Range-Aware Aggregation operator $\mathcal{A}_{row}(\cdot)$ that compresses the width dimension in $\mathbf{F}_{dec}$ to obtain a global motion descriptor $\mathbf{v}_y$ for each scan line $y$:
\begin{equation} 
    \mathbf{v}_y = \mathcal{A}_{row}(\mathbf F_{dec}) = \frac{1}{W} \sum_{x=1}^{W} \mathbf F_{dec}(x,y) \in \mathbb{R}^{64}.
\end{equation}

This operation enforces that all pixels within the same scan line share a unified motion representation, effectively suppressing local ambiguities and noise while preserving global deformation trends. Each row-wise descriptor $\mathbf{v}_y$ is then passed to a lightweight regression head $\mathcal{H}_{reg}$ to predict a 4-dimensional physical parameter vector:
\begin{equation}
    \mathbf{p}_y = \mathcal{H}_{reg}(\mathbf{v}_y)
    = \big[ k_x(y), k_y(y), b_x(y), b_y(y) \big]^T ,
\end{equation}
here, $k_x(y)$ and $k_y(y)$ denote the range-dependent slope coefficients, while $b_x(y)$ and $b_y(y)$ represent the corresponding intercept terms. Together, they define a 1D affine transformation that governs the geometric deformation of scan line $y$.

\paragraph{Parametric-to-Dense Flow Projection} 

Given the predicted physical parameters, the dense displacement field is reconstructed through a fixed projection layer. For a pixel located at $(x,y)$, the displacement vector $\mathbf{u}(x,y)$ is computed as:
\begin{equation}
    \mathbf{u}(x,y)
    = x \cdot
    \begin{bmatrix} k_x(y) \ k_y(y) \end{bmatrix}
    +
    \begin{bmatrix} b_x(y) \ b_y(y) \end{bmatrix}.
\end{equation}

The slope terms in this formulation model range-dependent geometric deviations (rotation), while the intercept terms capture uniform translational offsets. As a result, the predicted flow field is strictly confined to a low-dimensional subspace of physically plausible transformations. To perform geometric correction, the absolute sampling grid $\boldsymbol{\Phi}$ is constructed by adding the estimated displacement field $\mathbf{u}$ to the canonical identity grid $\mathbf{G}_{id}$:
\begin{equation}
    \boldsymbol{\Phi}(x,y) = \mathbf{G}_{id}(x,y) + \mathbf{u}(x,y)
    = (x + \Delta x, y + \Delta y).
\end{equation}

This grid defines the final coordinate mapping and is used by a differentiable sampler to warp the input SSS image, producing the geometrically corrected output.

\subsection{Knowledge Distillation via Hallucination}

Building upon the physically motivated deformation estimation, this part introduces a knowledge distillation mechanism that exploits privileged geometric information during training to enhance blind deformation inference. A Teacher network extracts the geometry-aware feature difference from paired data, while a Student network learns to approximate this reasoning from the distorted input through hallucinated context. Through multi-level distillation, geometric knowledge is transferred across representations and deformation outputs, enabling robust blind correction without $I_f$ at inference time.

\subsubsection{Privileged Feature Extraction (Teacher)}

During training, the Teacher network extracts geometric difference from paired samples $\{I_m, I_f\}$ through a shared encoder. This produces multi-scale representations $\mathbf{F}_{enc}^{m}$ and $\mathbf{F}_{enc}^{f}$ within a common feature space, where their difference directly serves as a characterization for geometric misalignment. Therefore, the privileged geometric context is defined as:
\begin{equation} 
    \mathbf{F}_{diff} = \mathbf{F}_{enc}^{m} - \mathbf{F}_{enc}^{f} \in \mathbb{R}^{\frac{H}{16} \times \frac{W}{16} \times 512},
\end{equation}
where $\mathbf{F}_{diff}$ denotes a latent correction signal capturing how distorted features should be transformed toward the corrected domain, and serves as geometry-aware supervision.

\subsubsection{Hallucination Context Module (HCM)}

\begin{figure}
	\centering
	\includegraphics[width=1\columnwidth]{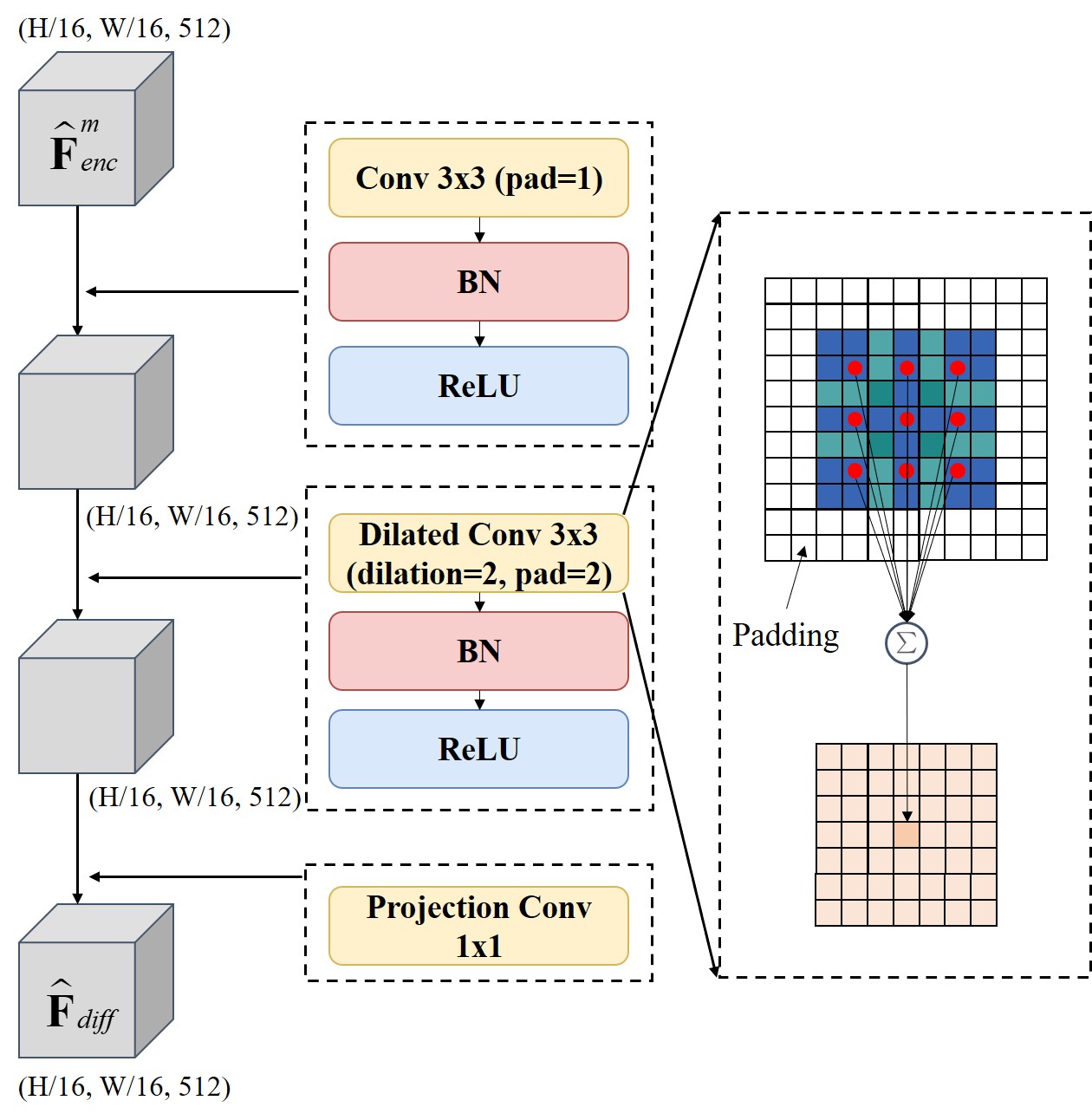}
    \caption{Structure of the Hallucination Context Module (HCM). Given distorted encoder features $\hat{\mathbf{F}}_{enc}^m$, the HCM predicts a hallucinated geometric difference $\hat{\mathbf{F}}_{diff}$ through three stages: local conditioning with a $3\times3$ convolution, global context aggregation using a dilated $3\times3$ convolution, and difference projection through a $1\times1$ convolution.  A key component is the Dilated Convolution (right side), which expands the receptive field to capture long-range, non-local distortion dependencies.}
	\label{fig:context}
\end{figure}	

Since the reference feature $\mathbf{F}_{enc}^{f}$ is unavailable during blind inference, the Student network cannot compute $\mathbf{F}_{diff}$ directly. To address this, we introduce the Hallucination Context Module (HCM), a learnable mapping $\mathcal{H}(\cdot)$ that infers an implicit geometric correction signal only from the distorted features. Formally, the HCM predicts a hallucinated geometric context as:
\begin{equation}
   \hat{\mathbf{F}}_{diff} = \mathcal{H}(\hat{\mathbf{F}}_{enc}^{m}),
\end{equation}
which is trained to approximate the privileged difference representation $\mathbf{F}_{diff}$ extracted by the Teacher.

Unlike conventional distillation schemes that aim to replicate static feature activations, the HCM is designed to capture the implicit relationship between distortion patterns and their corresponding geometric corrections. As illustrated in Fig. \ref{fig:context}, the HCM accounts for the non-local characteristics of SSS motion-induced distortions, such as long-range bending caused by yaw motion. To this end, it operates in three stages:

\begin{itemize}
    \item Local Conditioning: A $3 \times 3$ convolution refines $\hat{\mathbf{F}}_{enc}^{m}$ by suppressing noise and stabilizing local structural features.

    \item Global Context Aggregation: A dilated convolution with dilation rate $2$ expands the receptive field, enabling the module to aggregate long-range contextual information that is critical for modeling row-wise correlated deformations in SSS imagery.

    \item Difference Projection: A $1 \times 1$ convolution maps the aggregated features into the difference space, producing the hallucinated context $\hat{\mathbf{F}}_{diff}$.
\end{itemize}

While the encoder employs IN to  handle intensity variations, the HCM adopts Batch Normalization (BN). Since $\hat{\mathbf{F}}_{diff}$ represents relative geometric offsets rather than absolute intensities, its distribution is consistent across samples, making BN more suitable for convergence.

\subsubsection{Multi-Level Distillation Strategy}

To effectively transfer the Teacher’s geometry-aware inference behavior to the Student, we impose consistency constraints at three complementary levels:

\begin{itemize}

    \item \textbf{Feature Distillation ($\mathcal{L}_{enc}^{dis}$):} This term aligns the Student encoder output with the Teacher’s feature representation for the distorted input, ensuring a stable input to the HCM:
    \begin{equation}
        \mathcal{L}_{enc}^{dis} = \left| \hat{\mathbf{F}}_{enc}^{m} - \mathbf{F}_{enc}^{m} \right|_2^2, 
    \end{equation}
    where $\mathbf{F}_{enc}^{m}$ and $\hat{\mathbf{F}}_{enc}^{m}$ denote the Teacher and Student encoder features extracted from $I_m$, respectively.
    
    \item \textbf{Context Distillation ($\mathcal{L}_{ctx}^{dis}$:} As the core objective of the proposed framework, this term enforces the hallucinated geometric context to approximate the privileged difference representation:
    \begin{equation}
        \mathcal{L}_{ctx}^{dis} = \left\| \hat{\mathbf{F}}_{diff} - \mathbf{F}_{diff} \right\|_2^2.
    \end{equation}

    \item \textbf{Flow Distillation ($\mathcal{L}_{flow}^{dis}$):}  
    Finally, we align the deformation fields predicted by the Student $\boldsymbol{\Phi}_{S}$ and Teacher $\boldsymbol{\Phi}_{T}$ to ensure end-to-end functional consistency:
    \begin{equation}
        \mathcal{L}_{flow}^{dis} = \left\| \boldsymbol{\Phi}_{S} - \boldsymbol{\Phi}_{T} \right\|_2^2.
    \end{equation}

\end{itemize}

\subsection{Differentiable Forward Warping}

\begin{figure*}[htbp]
	\centering
	\includegraphics[width=1\linewidth]{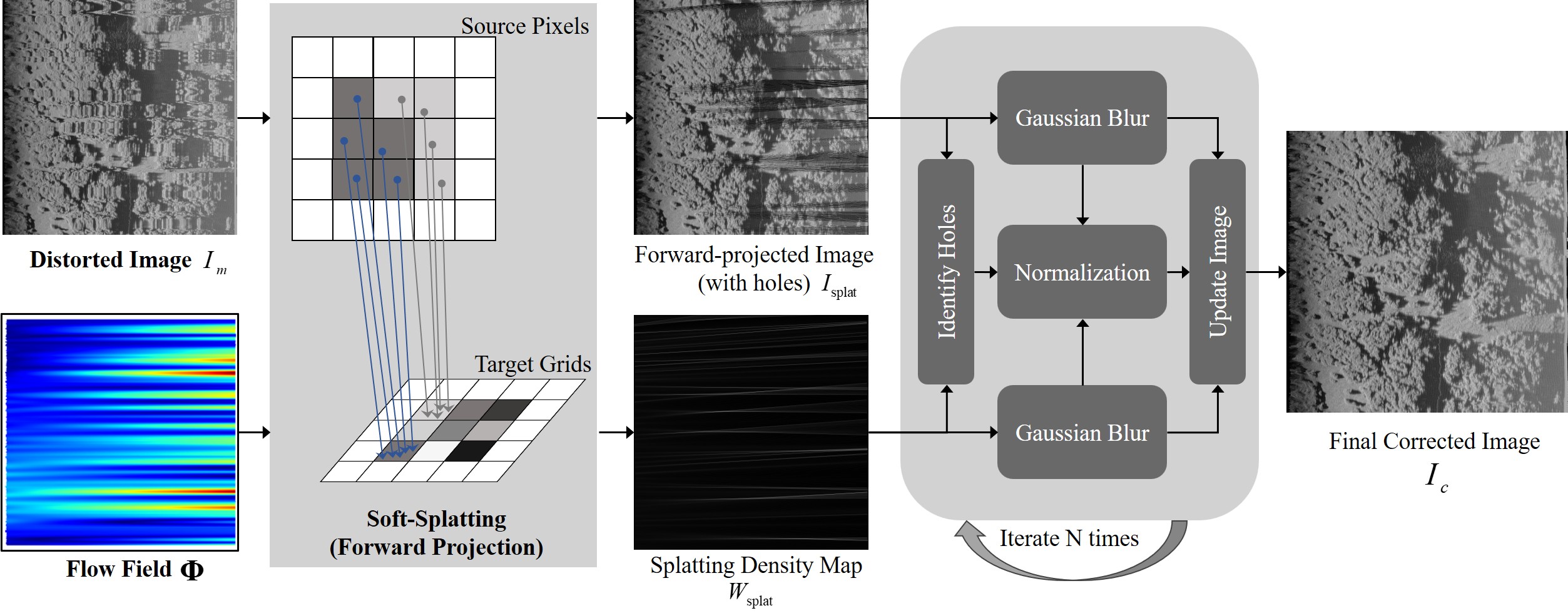}
	\caption{The Differentiable Forward Warping framework. Given the estimated deformation flow $\boldsymbol{\Phi}$, source pixels in the distorted image $I_m$ are forward-projected to the target grid through soft splatting, producing a forward-projected image and a splatting density map that shows unsampled regions (holes). An iterative hole-filling process based on Gaussian smoothing and normalization is then applied to diffuse valid measurements into holes, reconstructing the final corrected image $I_c$.}
	\label{fig:warping}
\end{figure*}

Given the estimated physically motivated deformation field $\mathbf{\Phi}$, the corrected SSS image is obtained by warping the distorted input $I_m$. Since the target image is defined on a geographically regularized domain rather than the original acquisition grid, the mapping from distorted scan samples to corrected coordinates is generally non-bijective, potentially producing overlaps (multiple measurements mapped to the same location) and holes (unsampled regions) after reprojection. This property is intrinsic to SSS geometric correction and makes standard backward sampling unsuitable, as it assumes that each target pixel has a valid source correspondence. We therefore adopt differentiable forward warping to project distorted scan samples onto the corrected domain. As illustrated in Fig. \ref{fig:warping}, source pixels are projected according to $\mathbf{\Phi}$ through soft splatting, followed by iterative hole filling, enabling geometrically consistent reconstruction and end-to-end optimization.

\subsubsection{Forward Projection via Soft-Splatting}

Given the estimated deformation flow field $\mathbf{\Phi}$, the forward warping process maps the distorted SSS image $I_m$ into the corrected domain. Let a discrete source pixel in $I_m$ be indexed by $\mathbf{x}=(x,y)$, where $x\in{0,\dots,W-1}$ and $y\in{0,\dots,H-1}$. The deformation field $\mathbf{\Phi}$ assigns each source pixel a continuous target location:
\begin{equation}
    \mathbf{\Phi}(\mathbf{x}) = (\Phi_x(\mathbf{x}), \Phi_y(\mathbf{x})) =
    \bigl(x + \Delta x(x,y), y + \Delta     y(x,y)\bigr),
\end{equation}
where $(\Delta x,\Delta y)$ denotes the estimated displacement.

The corrected image is defined on a discrete target grid, whose pixel indices are denoted by $\mathbf{x}'=(x',y')$, with $x'\in{0,\dots,W-1}$ and $y'\in{0,\dots,H-1}$. Since the continuous location $\mathbf{\Phi}(\mathbf{x})$ generally does not coincide with integer grid coordinates, the intensity value $I_m(\mathbf{x})$ is distributed to neighboring target pixels $\mathbf{x}'$ through a differentiable bilinear splatting kernel:
\begin{equation}
    \mathcal{K}(\mathbf{x}', \mathbf{\Phi}(\mathbf{x})) =
    \bigl[1-|x'-\Phi_x(\mathbf{x})|\bigr]_+ ,
\bigl[1-|y'-\Phi_y(\mathbf{x})|\bigr]_+,
\end{equation}
where $[\cdot]_+=\max(0,\cdot)$. Using average splatting \cite{A29}, the forward-projected image is computed as:
\begin{equation} 
    I_{\text{splat}}(\mathbf{x}') = \frac{ \sum_{\mathbf{x}} I_m(\mathbf{x}) \cdot \mathcal{K}(\mathbf{x}', \mathbf{\Phi}(\mathbf{x})) }{ \sum_{\mathbf{x}} \mathcal{K}(\mathbf{x}', \mathbf{\Phi}(\mathbf{x})) + \epsilon }, \label{eq:avg_splat} 
\end{equation}
where $\epsilon$ is a small constant for numerical stability. The denominator in Eq. \eqref{eq:avg_splat} defines a splatting density map:
\begin{equation}
    W_{\text{splat}}(\mathbf{x}') =
    \sum_{\mathbf{x}} \mathcal{K}(\mathbf{x}', \mathbf{\Phi}(\mathbf{x})),
\end{equation}
which measures the accumulated contribution from $I_m$. Target locations with $W_{\text{splat}}(\mathbf{x}') \approx 0$ correspond to regions not covered by any forward-projected measurements and are treated as holes in the corrected image.

\subsubsection{Iterative Hole Filling}

The forward-projected image $I_{\text{splat}}$ may contain holes caused by sparse forward mapping. To obtain a complete corrected image while preserving differentiability, we perform hole filling based on iterative normalized convolution, guided by the splatting density map. Specifically, a binary hole mask is defined as:
\begin{equation}
    \mathbf{M}(\mathbf{x}') =
    \mathbb{I}\left(W_{\text{splat}}(\mathbf{x}') \le \epsilon\right),
\end{equation}
where $\mathbb{I}(\cdot)$ denotes the indicator function.

We initialize the process with $I^{(0)} = I_{\text{splat}}$ and $W^{(0)} = W_{\text{splat}}$. At each iteration $i=1,\dots,N$, valid image information is diffused into hole regions through a sequence of weighted smoothing and normalization operations. First, the current image estimate and its density map are smoothed using a fixed Gaussian kernel $\mathcal{G}_\sigma$:
\begin{equation}
    \tilde{I}^{(i)} = (I^{(i-1)} \odot W^{(i-1)}) * \mathcal{G}_\sigma,
    \tilde{W}^{(i)} = W^{(i-1)} * \mathcal{G}_\sigma,
\end{equation}
where $\odot$ and $*$ denote element-wise multiplication and 2-D convolution, respectively. A locally normalized estimate is then obtained as:
\begin{equation}
    I_{\text{avg}}^{(i)} =
    \frac{\tilde{I}^{(i)}}{\tilde{W}^{(i)} + \epsilon}.
\end{equation}

The image is updated only at hole locations, while valid forward-projected pixels are preserved:
\begin{equation}
    I^{(i)} =
    (1-\mathbf{M}) \odot I^{(i-1)} +
    \mathbf{M} \odot I_{\text{avg}}^{(i)},
\end{equation}
which represents the weighted average of neighboring valid measurements. To ensure progressive expansion of valid support, the density map is updated accordingly:
\begin{equation}
    W^{(i)} =
    W^{(i-1)} + \mathbf{M} \odot \tilde{W}^{(i)}.
\end{equation}

After $N$ iterations (with $N=3$ in our experiments), the final geometrically corrected image is obtained as:
\begin{equation}
    I_c = I^{(N)}.
\end{equation}

\subsection{Joint Optimization Objectives}

\begin{table*}[t]
	\centering
	\caption{Summary of the three side-scan sonar datasets used in this paper.}
	\label{datasets}
	\renewcommand{\arraystretch}{1.2} % Increase row height for readability
	\setlength{\tabcolsep}{25pt}       % Adjusted to a normal spacing
	
	\begin{tabular}{lccc}
		\toprule
		\textbf{Dataset} & \textbf{I} & \textbf{II} & \textbf{III} \\
		\midrule
		Environment / Site & Sector N08 \cite{A30} & Sector N08 \cite{A30} & Sector N07 \cite{A30} \\
		Sonar Model & Marine Sonic Arc Scout MK II & Klein 3000H & Klein 3000H \\
		Operating Frequency & 900 kHz & 500 kHz & 500 kHz \\
		Sensor Range & 30-80 m & 80 m & 50-100 m \\
		Acquisition Altitude & $\sim$10\% of range &  $\sim$10\% of range &  $\sim$10\% of range \\
		Precise Navigation Data & Yes & No & No \\
		Geocoding available? & Yes & No & No \\
		Role & Training + Evaluation & Generalization Test & Generalization Test \\
		\bottomrule
	\end{tabular}
\end{table*}

\begin{table*}[t]
	\centering
	\caption{Quantitative comparison with state-of-the-art methods on the test set.}
	\vspace{1mm}
	\textbf{Note:} The symbols $\uparrow$ and $\downarrow$ indicate that higher and lower values correspond to better performance, respectively.
	\label{tab:sota}
	\renewcommand{\arraystretch}{1.2} % Increase row height for readability
	\setlength{\tabcolsep}{31pt}      % Adjust column spacing
	
	\begin{tabular}{lcccc}
		\toprule
		\textbf{Method} & \textbf{LPIPS} $\downarrow$ & \textbf{PSNR} $\uparrow$ & \textbf{MS-SSIM} $\uparrow$ & \textbf{NCC} $\uparrow$ \\
		\midrule
		DocUNet \cite{A24} & 0.444 & 21.477 & 0.515 & 0.636 \\
		GeoProj \cite{A38} & 0.301 & 24.481 & 0.616 & 0.826 \\
		TPS-STN \cite{A37} & 0.328 & 23.085 & 0.645 & 0.820 \\
		CycleGAN \cite{A35} & 0.302 & 24.613 & 0.627 & 0.822 \\
		U-Shape Baseline \cite{A34} & 0.271 & 25.362 & 0.723 & 0.876 \\
		VoxelMorph \cite{A36} & 0.170 & 25.218 & 0.754 & 0.830 \\
		\midrule
		\textbf{Ours} & \textbf{0.156} & \textbf{26.864} & \textbf{0.833} & \textbf{0.911} \\
		\bottomrule
	\end{tabular}
\end{table*}

The proposed physically motivated knowledge distillation framework is optimized by jointly performing geometric correction and transferring geometry-aware inference capability. To this end, training is driven by two complementary signals: direct geometric constraints derived from paired corrected references, and multi-level distillation constraints that guide the Student network to imitate the Teacher’s privileged geometric reasoning.

\subsubsection{Geometric Correction Losses}

Geometric correction losses are used to supervise the deformation field and the resulting corrected image.

\paragraph{Image Reconstruction Loss} 

Given the predicted deformation field $\mathbf{\Phi}$ and the corrected image $I_c$, we employ a hybrid reconstruction loss to ensure robust geometric evaluation against noise and intensity variations:
\begin{equation}
\mathcal{L}_{rec}
=
\lambda_{1} \left\| I_c - I_f \right\|_1
+
\lambda_{\text{perc}} \mathcal{L}_{\text{perc}}
+
\lambda_{\text{MI}} \mathcal{L}_{\text{MI}},
\end{equation}
where $\mathcal{L}_{\text{perc}}$ and $\mathcal{L}_{\text{MI}}$ denote the MobileNetV3-based perceptual loss and the intensity-invariant Mutual Information loss, respectively. The scalars $\lambda_{1}$, $\lambda_{\text{perc}}$, and $\lambda_{\text{MI}}$ are hyperparameters balancing these terms.

\paragraph{Flow Supervision Loss} 

When navigation priors permit establishing a geocoded ground truth flow $\mathbf{\Phi}_{gt}$, we constrain the predicted flow $\mathbf{\Phi}$ using the mean squared error:
\begin{equation}
\mathcal{L}_{flow}
=
\left| \mathbf{\Phi} - \mathbf{\Phi}_{gt} \right|_2^2,
\end{equation}
thereby enforcing precise geometric alignment between the distorted and corrected domains.

\paragraph{Smoothness Regularization}

While the parametric decoder imposes row-wise affine constraints, we further enforce spatial smoothness to suppress residual artifacts:
\begin{equation}
\mathcal{L}_{smt}
=
\left| \nabla^{2} \mathbf{\Phi} \right|_1,
\end{equation}
where $\nabla^{2}$ denotes the discrete Laplacian operator.

\subsubsection{Knowledge Distillation Losses}

To enable blind inference, we transfer the Teacher's privileged geometric reasoning to the Student through a multi-level distillation strategy that enforces consistency across feature, context, and flow levels:
\begin{equation}
\mathcal{L}_{dis} =  \lambda_{d1}\mathcal{L}_{enc}^{dis} + \lambda_{d2}\mathcal{L}_{ctx}^{dis} + \lambda_{d3}\mathcal{L}_{flow}^{dis},
\end{equation}
where $\lambda_{d1}$, $\lambda_{d2}$, and $\lambda_{d3}$ are non-negative hyperparameters balancing the distillation strength at the encoder, context hallucination, and deformation stages, respectively.

\subsubsection{Overall Optimization}

The physically motivated knowledge distillation framework employs tailored objectives for the Teacher and Student networks. The Teacher serves as a privileged geometric baseline, optimized through a combination of reconstruction, flow supervision, and smoothness constraints:
\begin{equation}
    \mathcal{L}_{Teacher}
    =
    \mathcal{L}_{rec}
    +
    \lambda_{f}\mathcal{L}_{flow}
    +
    \lambda_{s} \mathcal{L}_{smt},
\end{equation}
where $\lambda_{f}$ and $\lambda_{s}$ are weighting coefficients governing the influence of flow supervision and smoothness regularization, respectively.

To transfer this capability, the Student network is trained using the same geometric objectives augmented by the knowledge distillation loss: \begin{equation} 
    \mathcal{L}_{Student} = \mathcal{L}_{Teacher} + \mathcal{L}_{dis},\end{equation} 
which ensures that the Student not only adheres to physical constraints but also imitates the Teacher’s privileged reasoning, enabling accurate geometric correction during blind inference.

\section{Experiment}

\begin{figure*}[htbp]
	\centering
	\includegraphics[width=1\linewidth]{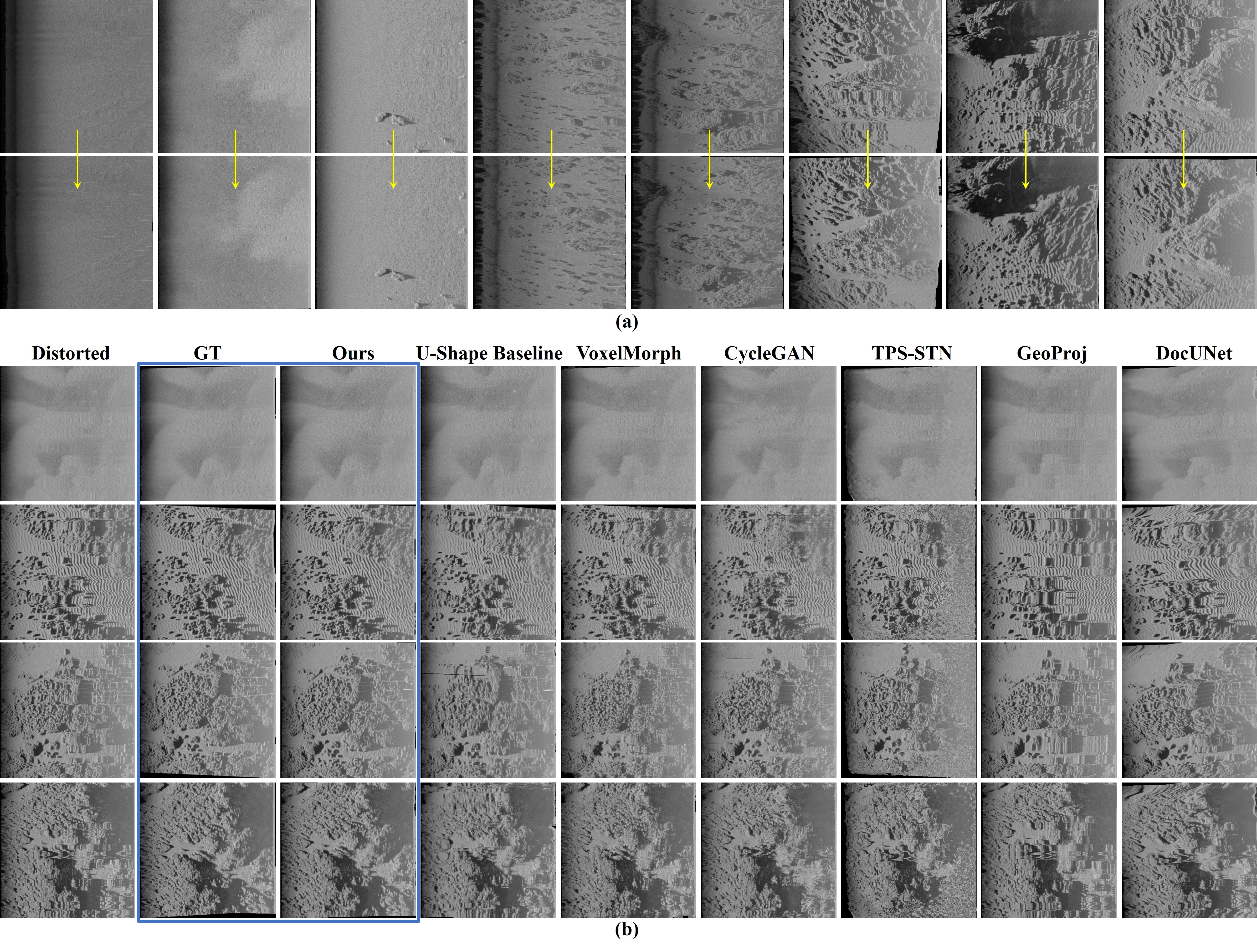}
	\caption{Visualization evaluation and comparative analysis.
		(a) Correction results under diverse seabed conditions. From left to right, examples include sparse textures, sand ripples, and complex rocky terrains. The proposed method effectively removes motion-induced distortions while preserving target shapes, ripple continuity, and structural consistency across different scenarios. (b) Visual comparison with state-of-the-art methods. Direct regression models (U-Shape Baseline, VoxelMorph) suffer from texture blurring, CycleGAN introduces unrealistic structures, TPS-STN fails to capture row-wise deformations, and cross-domain methods (GeoProj, DocUNet) produce severe tearing. In contrast, as highlighted by the blue box, our method achieves the highest consistency with the Ground Truth (GT). Note that the GT is obtained through navigation-based correction, which performs geocoding relying on high-precision navigation and attitude information.}
	\label{fig6}
\end{figure*}

\begin{table*}[t]
	\centering
	\caption{Ablation study quantitatively analyzing the contribution of each component in the proposed PCKD framework.}
	\vspace{1mm}
	\textbf{Note:} The checkmark ($\checkmark$) and cross ($\times$) symbols denote the inclusion and exclusion of specific modules or loss terms, respectively. The Decoder column indicates whether the Parametric (physically motivated) or Dense (unconstrained) decoder was employed. 
	
	\label{tab:ablation_full}
	\renewcommand{\arraystretch}{1.2}
	\setlength{\tabcolsep}{10pt}     
	
	\begin{tabular}{l|ccccc|cccc}
		\toprule
		\multirow{2}{*}{\textbf{Method}} & \multicolumn{5}{c|}{\textbf{Experimental Settings}} & \multicolumn{4}{c}{\textbf{Quantitative Results}} \\
		\cmidrule(lr){2-6} \cmidrule(lr){7-10}
		& $\mathcal{L}_{flow}$ & $\mathcal{L}_{flow}^{dis}$ & $\mathcal{L}_{enc}^{dis}$ & $\mathcal{L}_{ctx}^{dis}$ & \textbf{Decoder} & \textbf{LPIPS} $\downarrow$ & \textbf{PSNR} $\uparrow$ & \textbf{MS-SSIM} $\uparrow$ & \textbf{NCC} $\uparrow$ \\
		\midrule
		\textbf{Ours} & \checkmark & \checkmark & \checkmark & \checkmark & Parametric & \textbf{0.156} & \textbf{26.864} & \textbf{0.833} & \textbf{0.911} \\
		\midrule
		w/o Teacher Framework & \checkmark & $\times$ & $\times$ & $\times$ & Parametric & 0.206 & 25.458 & 0.722 & 0.867 \\
		w/o GT Flow & $\times$ & \checkmark & \checkmark & \checkmark & Parametric & 0.288 & 25.637 & 0.737 & 0.880 \\
		w/o Physics Constraint & \checkmark & \checkmark & \checkmark & \checkmark & \textbf{Dense} & 0.170 & 26.212 & 0.803 & 0.896 \\
		w/o Flow Distill & \checkmark & $\times$ & \checkmark & \checkmark & Parametric & 0.168 & 26.435 & 0.810 & 0.900 \\
		w/o Feature Distill & \checkmark & \checkmark & $\times$ & \checkmark & Parametric & 0.170 & 26.336 & 0.802 & 0.896 \\
		w/o Context Distill & \checkmark & \checkmark & \checkmark & $\times$ & Parametric & 0.172 & 26.105 & 0.793 & 0.894 \\
		\bottomrule
	\end{tabular}
	
\end{table*}

\begin{figure*}[htbp]
	\centering
	\includegraphics[width=1\linewidth]{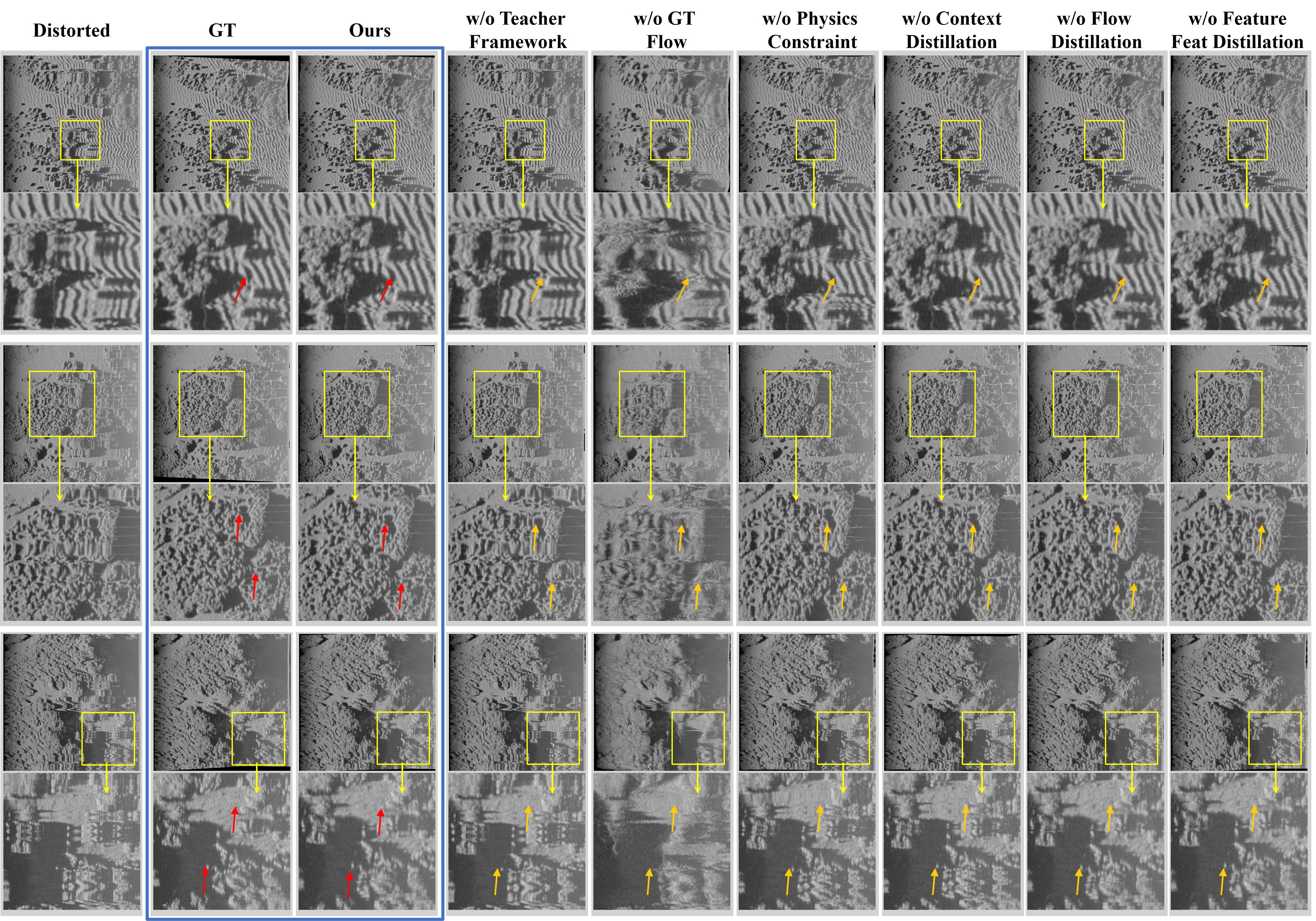}
	\caption{Visual ablation study of the proposed framework. The columns compare the distorted input, GT, and our full method against variants lacking specific components, including the Teacher framework, ground-truth flow supervision, physical constraints, and individual distillation terms. Yellow boxes indicate enlarged regions, and arrows highlight texture orientations and local distortions.  As highlighted by the blue box, the proposed full model recovers geometry and texture consistent with the GT (red arrows). Removing the Teacher or flow supervision results in incomplete correction and blurred structures, while disabling the physical constraint introduces local jitter. }
	\label{fig7}
\end{figure*}

\subsection{Data Description}

We employ three SSS datasets acquired from different environments and sonar platforms. The detailed acquisition parameters and sonar specifications for all three datasets are summarized in Table \ref{datasets}. Dataset I is used for training and quantitative evaluation, as it is accompanied by high-precision navigation and attitude measurements, which enable reliable geocoding of the raw sonar observations. This geocoding process produces geometrically corrected reference images and the corresponding deformation fields, thereby providing samples required for Teacher–Student training and objective performance evaluation. In contrast, Datasets II and III lack accurate navigation data to support geocoding-based correction and are therefore reserved for assessing the generalization capability of the proposed blind geometric correction method under unseen conditions.

\subsection{Evaluation Metrics}

Let $I_f$ and $I_c$ denote the reference image and the corrected output, respectively, both normalized to the range $[0,1]$. We employ four complementary metrics to assess geometric correction quality from pixel-level accuracy to perceptual consistency.

\subsubsection{Peak Signal-to-Noise Ratio (PSNR)}

PSNR measures reconstruction accuracy by quantifying the inverse of the mean squared error between $I_c$ and $I_f$:
\begin{equation}
    \text{PSNR}
    =
    10 \log_{10}
    \left(
    \frac{1}{\frac{1}{|\Omega|}\sum_{x \in \Omega}
    \left( I_c(x) - I_f(x) \right)^2}
    \right),
\end{equation}
where $\Omega$ denotes the image domain and $x$ indexes spatial locations. Higher PSNR values indicate smaller overall reconstruction error.

\subsubsection{Multi-Scale Structural Similarity (MS-SSIM)}

MS-SSIM \cite{A32} evaluates structural consistency between the corrected output $I_c$ and the geocoded reference $I_f$ across multiple spatial resolutions:
\begin{equation}
\text{MS-SSIM}=\left[l_M(I_c, I_f)\right]^{\alpha_M}\prod_{j=1}^{M}\left[c_j(I_c, I_f)\right]^{\beta_j}\left[s_j(I_c, I_f)\right]^{\gamma_j},
\end{equation}where $l_M$ is the luminance similarity at scale $M$, $c_j$ and $s_j$ represent the contrast and structural terms at scale $j$, and $(\alpha_M, \beta_j, \gamma_j)$ follow the standard MS-SSIM weights. Higher values indicate better structural consistency.

\subsubsection{Normalized Cross-Correlation (NCC)}

NCC measures the linear correlation between $I_c$ and $I_f$, providing robustness to global intensity variations:
\begin{equation}
\text{NCC}
=
\frac{1}{|\Omega|\, \sigma_f \sigma_c}
\sum_{x \in \Omega}
\left( I_f(x) - \mu_f \right)
\left( I_c(x) - \mu_c \right),
\end{equation}
where $\mu_f$, $\mu_c$ and $\sigma_f$, $\sigma_c$ denote the mean and standard deviation of $I_f$ and $I_c$, respectively. An NCC value closer to $1$ indicates stronger alignment.

\subsubsection{Learned Perceptual Image Patch Similarity (LPIPS)}

LPIPS \cite{A33} evaluates perceptual similarity by comparing deep feature representations extracted from a pretrained network:
\begin{equation}
\text{LPIPS}
=
\sum_{l}
\frac{1}{H_l W_l}
\sum_{h,w}
\left\|
w_l \odot
\left(
\phi_f^{\,l}(h,w)
-
\phi_c^{\,l}(h,w)
\right)
\right\|_2^2,
\end{equation}
where $\phi_f^{\,l}$ and $\phi_c^{\,l}$ denote feature maps of $I_f$ and $I_c$ at layer $l$, $(H_l, W_l)$ are the spatial dimensions, and $w_l$ are learned channel-wise weights. Lower LPIPS values correspond to higher perceptual and semantic consistency.

\subsection{Experimental settings}

\textit{Experimental Environment Configuration:} All experiments were conducted on a high-performance workstation running Ubuntu 20.04.6 LTS. The hardware configuration consists of dual Intel Xeon Gold 6230 CPUs (2.10 GHz) paired with 376 GB of RAM. Deep learning acceleration is provided by two NVIDIA Quadro RTX 6000 GPUs, each with 24 GB of VRAM. The software environment is built upon PyTorch, leveraging CUDA 12.0 and NVIDIA driver 525.147 for efficient parallel computing.

% \subsubsection{Network Implementation Details}

% Implemented in PyTorch, our framework processes input images resized to a fixed resolution of $512 \times 512$. The Shared Encoder adopts a five-stage ResNet-style architecture (channel depths: $64, 128, 256, 512$). Notably, we employ Instance Normalization (IN) instead of Batch Normalization to preserve ping-wise acoustic characteristics. The Physically-Constrained Decoder utilizes a global average pooling head to regress four affine motion parameters ($k_u, b_u, k_v, b_v$) per row. For the Student network, the Hallucination Context Module incorporates dilated convolutions ($rate=2$) to expand the receptive field for global distortion perception.

\textit{Training Strategy:} All input images are resized to $512 \times 512$. Both the Teacher and Student networks are trained using the AdamW optimizer with an initial learning rate of $1 \times 10^{-4}$, a weight decay of $1 \times 10^{-4}$, and a cosine annealing scheduler. The Teacher is trained for 300 epochs using paired samples $\{I_m, I_f\}$ with a batch size of 8, where the reconstruction loss weights are set to $\lambda_{1}=1.0$, $\lambda_{\text{perc}}=0.1$, and $\lambda_{\text{MI}}=0.2$, and the flow and smoothness terms are weighted by $\lambda_{f}=2.0$ and $\lambda_{s}=0.1$, respectively. The Student is initialized from the Teacher’s encoder and trained for 400 epochs with the Teacher frozen, where the distillation weights are set to $\lambda_{d1}=0.5$, $\lambda_{d2}=1.0$, and $\lambda_{d3}=0.5$.

\subsection{Visualization results}

Fig. \ref{fig6}(a) presents the blind geometric correction performance across diverse seabed complexities. The top row displays the distorted inputs, while the bottom row shows the corrected outputs. In sparse texture regions (Cols. 1–3), the method corrects geometric distortions, restoring the real spatial shapes of isolated targets. For periodic textures like sand ripples (Cols. 4–5), motion-induced errors are effectively smoothed, producing continuous wave patterns consistent with reality. And in complex rocky terrains (Cols. 6–8), the physically motivated decoder preserves structural consistency, aligning acoustic shadows and texture details without introducing image tearing or resampling artifacts.

\subsection{Comparison with State-of-the-Art}

\begin{figure*}[htbp]
	\centering
	\includegraphics[width=1\textwidth]{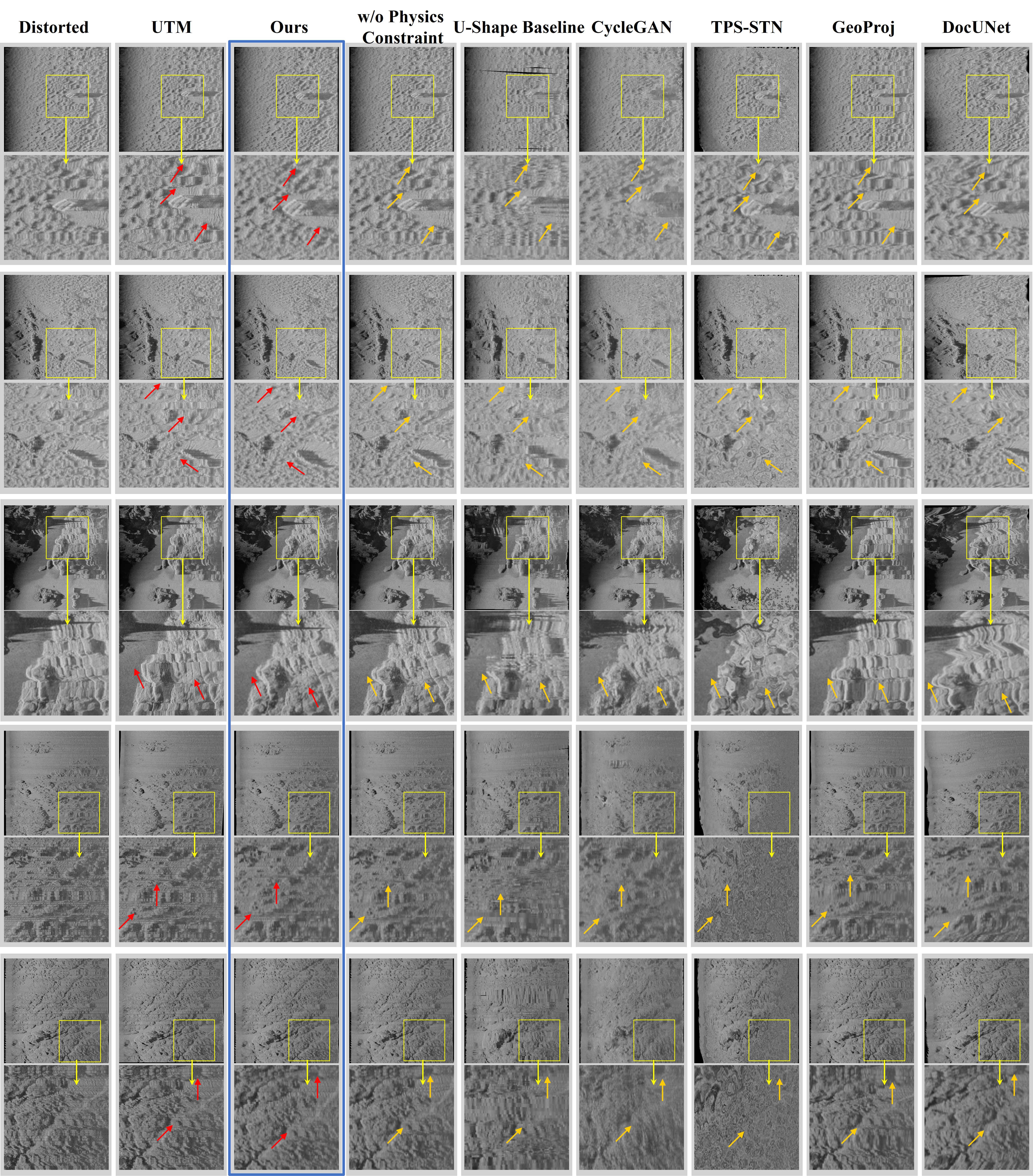}
	\caption{Generalization results on unseen datasets without fine-tuning. The first six rows correspond to Dataset II and the last four rows to Dataset III. The proposed method (blue box) shows superior zero-shot transferability, preserving more continuous geometric structures than UTM geocoding and other baselines (red arrows). In contrast, the w/o Physics Constraint variant (Column 4) exhibits structural discontinuities (orange arrows), highlighting the role of the physical prior in improving robustness under domain shifts. In these scenarios, UTM geocoding is severely degraded by unreliable navigation data and cannot be regarded as ground truth, while other baselines suffer from blurring or tearing artifacts due to the lack of explicit sonar geometric modeling.}
	\label{fig8}
\end{figure*}	

To the best of our knowledge, no deep learning method has been developed for geometric correction of SSS images. Therefore, besides general-purpose registration and transformation models, we also include representative deep models designed for related geometric correction tasks in other domains, such as optical image distortion correction and document image unwarping. Specifically, we compare the proposed framework with six representative baseline methods covering different modeling paradigms, including U-shaped encoder-decoder baseline \cite{A34} (direct regression), CycleGAN \cite{A35} (unpaired generative translation), VoxelMorph \cite{A36} (unsupervised dense registration), TPS-STN \cite{A37} (parametric spatial transformation), and two approaches adapted from related geometric tasks, GeoProj \cite{A38} (optical distortion correction) and DocUNet \cite{A24} (document unwarping). To ensure a fair comparison, all methods are retrained on the same SSS dataset under identical experimental settings.

\subsubsection{Quantitative Analysis}

Quantitative results in Table \ref{tab:sota} show that the proposed method (Ours) outperforms all baselines across all four metrics, with an LPIPS of 0.156, PSNR of 26.864 dB, MS-SSIM of 0.833, and NCC of 0.911. Among the baselines, VoxelMorph attains the second-best results in structural metrics (LPIPS and MS-SSIM), while U-Shape Baseline achieves the second-highest scores in intensity-based metrics (PSNR and NCC). This indicates that while standard registration (VoxelMorph) and direct regression (U-Shape) offer partial improvements, neither matches the comprehensive performance of our physically motivated approach. In contrast, GeoProj and DocUNet exhibit significantly degraded performance, confirming the limited transferability of optical and document-specific priors to sonar data.

\subsubsection{Visual Analysis}

Visual comparisons in Fig.~\ref{fig6}(b) further validate these findings. Direct regression methods (U-Shape Baseline, VoxelMorph) correct global deformation but suffer from significant textural blurring. CycleGAN generates spurious textures that deviate from reality due to the lack of geometric alignment. TPS-STN imposes an erroneous global smoothness constraint through sparse control points, failing to model row-wise deformation. Furthermore, Cross-domain approaches (GeoProj, DocUNet) cannot model sonar mechanism, introducing severe pixel tearing rather than effective correction. In contrast, the proposed method eliminates motion distortions while preserving sharp textures, restoring the continuity of sand ripples and the coherence of complex rocks to closely resemble the Ground Truth.

\subsection{Ablation Study}

To isolate the contribution of each component, we conduct an ablation study (Table \ref{tab:ablation_full}) by selectively disabling key modules. The variants include removing the Teacher framework or ground-truth flow supervision, replacing the physically motivated parametric decoder with a dense decoder, and disabling individual distillation terms ($\mathcal{L}_{enc}^{dis}$, $\mathcal{L}_{ctx}^{dis}$, $\mathcal{L}_{flow}^{dis}$). All models are trained under the same experimental settings for fair comparison.

\subsubsection{Quantitative Analysis}

Quantitative results in Table \ref{tab:ablation_full} demonstrate that the full model achieves the best performance. Removing the Teacher framework leads to the most severe structural degradation (MS-SSIM: $0.833 \rightarrow 0.722$), confirming the vital role of privileged geometric guidance. Meanwhile, disabling ground-truth flow supervision results in the worst perceptual quality (LPIPS: 0.288), indicating that explicit geometric information remains critical for stable learning. Substituting the parametric decoder with a dense model reduces both MS-SSIM and NCC, validating the benefit of physical constraints. Among distillation components, removing context distillation causes the largest drop ($0.793$ in MS-SSIM), proving it is the most critical transfer module compared to feature and flow distillation.

\subsubsection{Visual Analysis}

Visual comparisons in Fig. \ref{fig7} further validate these contributions. The figure displays both global results and magnified areas (yellow boxes) to highlight fine details, with arrows indicating texture orientations. Removing the Teacher framework leads to incomplete geometric correction, where stripe-like textures remain significantly distorted (orange arrows). Without ground-truth flow supervision, the results suffer from structural blurring and reduced clarity. Disabling the physical constraint introduces high-frequency jitter and local irregularities in the magnified regions. In contrast, the full model produces smoother and more coherent corrections, while ablations of distillation terms show slight degradation in fine-scale consistency.

\subsection{Generalization Study}

To evaluate the generalization capability of the proposed framework, we conduct experiments on Datasets II and III, where accurate navigation and attitude data is unavailable. The model trained on Dataset I is directly applied to these unseen datasets without fine-tuning, enabling a zero-shot evaluation under domain shifts. Since geocoded references are absent, the assessment relies on visual analysis, as shown in Fig. \ref{fig8}, where the first six rows correspond to Dataset II and the last four rows to Dataset III. For completeness, we include both state-of-the-art baselines and a w/o Physics Constraint variant to analyze the factors affecting generalization behavior.

\subsubsection{Impact of Physical Constraints} 

A direct comparison between the proposed method and the w/o Physics Constraint variant highlights the critical role of geometric priors.  As observed across multiple rows in Fig. \ref{fig8}, although the unconstrained dense decoder performs competitively on the source domain (Dataset I), it degrades on unseen datasets, exhibiting severe structural discontinuities. This contrast indicates that without geometric priors, the network overfits to specific textures. Conversely, our row-wise physical constraint compels the model to learn sonar motion laws rather than memorizing visual patterns, ensuring robust generalization.

\subsubsection{Comparison with State-of-the-Art}

Fig. \ref{fig8} presents a comparison against baselines on the unseen datasets. The result labeled as UTM corresponds to conventional navigation-based geocoding. Because the navigation data in these datasets are unreliable, the resulting geocoded images exhibit severe misalignment and therefore are used only as a reference baseline rather than as ground truth. Among learning-based methods, U-Shape Baseline suffers from noticeable blurring and residual jitter. CycleGAN generates fake textures inconsistent with the actual terrain, while TPS-STN produces overly smoothed results with severe stretching. Furthermore, cross-domain methods (GeoProj, DocUNet) exhibit severe degradation with global pixel tearing, confirming the incompatibility of optical priors with acoustic data. In contrast, the proposed method consistently delivers geometrically stable and coherent corrections across both datasets, demonstrating superior generalization capability under blind conditions.

\section{Conclusion}

In this paper, we proposed a physically motivated knowledge distillation framework for blind geometric correction of SSS imagery, aiming to address the challenge of deformation estimation in the absence of navigation and attitude information. The proposed teacher–student paradigm enables the transfer of geometry-aware knowledge from a privileged training setting to a deployable blind student model, allowing geometric correction to be performed directly from a single distorted SSS image. Experimental results demonstrate that the proposed method outperforms representative state-of-the-art baselines across multiple quantitative metrics, achieving improved geometric consistency while avoiding non-physical artifacts commonly observed in unconstrained or cross-domain approaches. Ablation studies confirm that both the physically motivated parametric decoder and the knowledge distillation strategy are essential for stable and accurate blind correction. Furthermore, zero-shot evaluations on unseen datasets without reliable auxiliary data indicate that incorporating sonar-specific geometric priors leads to more stable and coherent geometric correction results compared to unconstrained alternatives. Despite these advantages, the proposed method remains limited by the representation capacity of the row-wise affine deformation model, which cannot fully describe complex seabed geometry, such as sharp terrain variations. Future work will therefore explore the integration of additional geometric information, such as bathymetric or height estimation, to better address terrain-induced geometric effects and further improve the applicability of physically motivated learning for practical SSS data processing.

\section*{Acknowledgments}

This work was supported in part by the Science, Technology and Innovation Commission of Shenzhen Municipality (Grant Nos. JCYJ20241202124931042, ZDCYKCX20250901093900002), and in part by the Spanish government through projects IURBI (CNS2023-144688) and ASSiST (PID2023-149413OB-I00).

Can Lei also gratefully acknowledges the scholarship support from the China Scholarship Council (CSC).

\bibliographystyle{IEEEtran}
\bibliography{refernew}

%{\appendices
%\section*{Proof of the First Zonklar Equation}
%Appendix one text goes here.
% You can choose not to have a title for an appendix if you want by leaving the argument blank
%\section*{Proof of the Second Zonklar Equation}

\vfill

\end{document}